\documentclass[pra,twocolumn,showpacs,amsmath,amssymb,floatfix]{revtex4}
\usepackage{graphicx}
\usepackage{dcolumn}
\usepackage{bm}

\begin{document}

\title{Derivation of Maxwell-Bloch-type equations by projection of quantum models}
\author{Hideo Mabuchi} \email{hmabuchi@stanford.edu}
\affiliation{Physical Measurement and Control, Edward L.\ Ginzton Laboratory, Stanford University}

\date{\today}

\begin{abstract}
A simple algebraic procedure is described for deriving Maxwell-Bloch-type equations from single-atom cavity quantum electrodynamics (cavity QED) master equations via orthogonal projection onto a manifold of semiclassical states. In particular the usual Maxwell-Bloch Equations are obtained---up to a state-dependent correction factor of order unity---straightforwardly from the unconditional Jaynes-Cummings master equation. The technique of projecting onto a semiclassical manifold can also be applied with conditional master equations (quantum filters), leading to stochastic simulation models that include multiplicative noise terms associated with fluctuations of the atomic dipole. The utility of such models is briefly explored in the context of single-atom absorptive bistability.
\end{abstract}

\maketitle

\noindent For many years, cavity quantum electrodynamics (cavity QED) in the optical regime has provided a canonical setting for theoretical and experimental investigations of non-equilibrium quantum statistical mechanics and the quantum-classical transition \cite{Lugi84,Berm94,Mabu02}. Historically much attention has been focused on the comparison of semi-classical models of cavity QED, exemplified by the Maxwell-Bloch Equations (MBE's) \cite{Lugi84}, and quantum-mechanical models such as the Jaynes-Cummings master equation \cite{Carm93}. It is now generally accepted that the MBE's are most directly applicable in scenarios where many intracavity atoms each contribute weakly to the overall cooperativity \cite{Lugi84,Carm86a,Rose91}, whereas quantum-mechanical models are required to account for experimental data in single-atom scenarios with strong coupling \cite{Hood98,Brun96}. There are however some subtleties to this distinction as, {\it e.g.}, it can be shown that MBE's govern the dynamics of mean values of quantum observables in single-atom cavity QED scenarios in the ``bad cavity limit'' with weak excitation \cite{Rice88}. Rigorous distinction between semi-classical and quantum phenomena in cavity QED is of fundamental interest for both the study of mesoscopic physics and quantum information science \cite{Mabu02}.

In investigations based on numerical simulation of quantum models \cite{Sava88,Kili91,Alsi91,Mabu98,Arme06}, it has been observed that the MBE's actually provide fairly accurate guidance when searching parameter space for bifurcation-like phenomena in single-atom cavity QED with strong driving fields. This extended relevance of the MBE's in regimes of strong coupling and nonlinear dynamics is somewhat surprising, and there has been no prior analysis of the quantitative relation between the MBE's and any fully quantum-mechanical model of single-atom cavity QED that would be valid under such conditions. In this article I point out that a simple technique, introduced recently into quantum optics by Van~Handel \cite{vanH05}, can be used to show that the MBE's essentially correspond (up to a factor of order unity) to an orthogonal projection of the single-atom cavity QED master equation onto an intuitive manifold of semiclassical states. No assumptions regarding parameter regime are required. Applying the same technique to stochastic master equations \cite{Carm93,Wise93,Goet94,Gard04} (quantum filters \cite{Bout06}) for detection of the atomic fluorescence \cite{Wise95,Bout04}, I arrive at modified MBE's with noise terms that constitute semi-classical cavity QED models incorporating atomic dipole fluctuations. These models provide an interesting complement to stochastic extensions of the MBE's that have been proposed for incorporating electromagnetic field fluctuations \cite{Carm86b}.

As usual we consider cavity QED models in which a two-level atom is coupled to a single mode of the electromagnetic field \cite{Jayn63,Carm93}. Our general strategy \cite{vanH05} for obtaining semi-classical MBE's is to project a quantum equation of motion, which we denote abstractly
\begin{equation}
d\theta_t = {\cal L}\left[\theta_t\right],\label{eq:qm}
\end{equation}
onto the manifold of states
\begin{eqnarray}
\rho%(p_r,p_i,D,x_r,x_i)
&\equiv&\frac{1}{2}\left(nI-\sqrt{2}p_r\sigma_x+\sqrt{2}p_i\sigma_y-D\sigma_z\right)
\otimes\vert\alpha\rangle\langle\alpha\vert,\nonumber\\
\alpha&\equiv&\frac{\gamma_\perp}{\sqrt{2}g_0}(x_r+ix_i),\label{eq:manif}
\end{eqnarray}
where $\vert\alpha\rangle$ is a coherent state with complex amplitude $\alpha$ for the electromagnetic field mode, $\gamma_\perp$ is the atomic dipole decay rate and $g_0$ is the ``vacuum Rabi frequency'' characterizing the strength of the atom-cavity coupling. In our parametrization $\sigma_{x,y,z}$ are the usual Pauli operators and $I$ represents the identity operator on the atomic Hilbert space. We are thus restricting the system dynamics to unentangled quantum states in which the field is semi-classical and the atom assumes an arbitrary mixed state. Our chosen parametrization of the atom-cavity density operator in terms of the real scalar variables $p_r$, $p_i$, $D$, $x_r$ and $x_i$ will lead us to Maxwell-Bloch-type equations. The extra parameter $n$ is included for technical reasons to allow us to enforce normalization of the density operator in the projected equations, and will drop out of the final equations.

Orthogonal projection of Eq.~(\ref{eq:qm}) is accomplished via the general rule
\begin{equation}
\Pi_{{\rm span}\{\nu_i\} }[x]=\sum_i\frac{\langle\nu_i,x\rangle\nu_i}{\langle\nu_i,\nu_i\rangle},\label{eq:proj}
\end{equation}
with $x\leftrightarrow d\theta_t$ and with $\{\nu_i\}$ corresponding to a full set of tangent vectors to the parametrized manifold of states:
\begin{equation}
\{\nu_i\}\leftrightarrow \left\{ \frac{\partial\rho}{\partial n},
\frac{\partial\rho}{\partial p_r},\frac{\partial\rho}{\partial p_i},
\frac{\partial\rho}{\partial D},
\frac{\partial\rho}{\partial x_r}, \frac{\partial\rho}{\partial x_i}\right\}.
\end{equation}
We choose the inner product $\langle X,Y\rangle={\rm Tr}[X^*Y]$ on the space of atom-cavity density operators, which will prove to be especially convenient because Pauli operators play a central role in our parametrization. Note that other choices of inner product could be made and might lead to somewhat different final equations.

After some straightforward calculations we find
\begin{equation}
\frac{\partial\rho}{\partial n}=\frac{1}{2}I\otimes\vert\alpha\rangle\langle\alpha\vert,\quad
\frac{\partial\rho}{\partial D}=-\frac{1}{2}\sigma_z\otimes\vert\alpha\rangle\langle\alpha\vert,
\end{equation}
\begin{equation}
\frac{\partial\rho}{\partial p_r}=-\frac{1}{\sqrt{2}}\sigma_x\otimes\vert\alpha\rangle\langle\alpha\vert,\quad
\frac{\partial\rho}{\partial p_i}=\frac{1}{\sqrt{2}}\sigma_y\otimes\vert\alpha\rangle\langle\alpha\vert,
\end{equation}
and
\begin{eqnarray}
\frac{\partial\rho}{\partial x_r}&=&%\frac{\gamma_\perp}{\sqrt{2}g_0}
\rho_{\rm at}\otimes\left[(a^*-\alpha^*)\vert\alpha\rangle\langle\alpha\vert
+\vert\alpha\rangle\langle\alpha\vert(a-\alpha)\right],\nonumber\\
\frac{\partial\rho}{\partial x_i}&=&%\frac{i\gamma_\perp}{\sqrt{2}g_0}
\rho_{\rm at}\otimes\left[i(a^*-\alpha^*)\vert\alpha\rangle\langle\alpha\vert
-i\vert\alpha\rangle\langle\alpha\vert(a-\alpha)\right],
\end{eqnarray}
where
\begin{equation}
\rho_{\rm at}\equiv\frac{1}{2}\left( nI-\sqrt{2}p_r\sigma_x+\sqrt{2}p_i\sigma_y-D\sigma_z\right).
\end{equation}
As given these tangent vectors are mutually orthogonal but the last two are not normalized:
\begin{eqnarray}
\left\langle\frac{\partial\rho}{\partial x_r},\frac{\partial\rho}{\partial x_r}\right\rangle
&=&\left\langle\frac{\partial\rho}{\partial x_i},\frac{\partial\rho}{\partial x_i}\right\rangle\nonumber\\
&=&n^2+2p_r^2+2p_i^2+D^2.
\end{eqnarray}
This provides the denominator required in Eq.~(\ref{eq:proj}).

Finally we simply write
\begin{eqnarray*}
d\rho_t&=&\Pi_{{\rm span}\{\nu_i\} }[d\theta_t]\\
&=&\frac{\partial\rho}{\partial n}dn+\frac{\partial\rho}{\partial p_r}dp_r+\frac{\partial\rho}{\partial p_i}dp_i\\
&&+\frac{\partial\rho}{\partial D}dD+\frac{\partial\rho}{\partial x_r}dx_r+\frac{\partial\rho}{\partial x_i}dx_i,
\end{eqnarray*}
and thus, using Eq.~(\ref{eq:proj}) and orthogonality of the tangent vectors,
\begin{equation}
dn=\left\langle\frac{\partial\rho}{\partial n},d\theta_t\right\rangle,\quad
dD=\left\langle\frac{\partial\rho}{\partial D},d\theta_t\right\rangle,
\end{equation}
\begin{equation}
dp_r=\left\langle\frac{\partial\rho}{\partial p_r},d\theta_t\right\rangle,\quad
dp_i=\left\langle\frac{\partial\rho}{\partial p_i},d\theta_t\right\rangle,
\end{equation}
and
\begin{eqnarray}
dx_r&=&(n^2+2p_r^2+2p_i^2+D^2)^{-1}\left\langle\frac{\partial\rho}{\partial x_r},d\theta_t\right\rangle,\quad
\nonumber\\
dx_i&=&(n^2+2p_r^2+2p_i^2+D^2)^{-1}\left\langle\frac{\partial\rho}{\partial x_i},d\theta_t\right\rangle.
\end{eqnarray}
All that remains in order to obtain concrete semi-classical equations of motion for the real scalar parameters is to insert a specific quantum model for ${\cal L}[\theta_t]$. It is interesting to note that, in any given time-step, the error associated with projection onto the target manifold can be quantified by computing the norm of $(d\theta_t-d\rho_t)$.

We first consider the unconditional Jaynes-Cummings master equation \cite{Carm93}, written in a rotating frame at the frequency $\omega_l$ of the external driving field ($\hbar=1)$:
\begin{eqnarray}
d\theta_t&=&-i[H,\theta_t]dt+\kappa(2a\theta_t a^* - a^*a\theta_t - \theta_ta^*a)dt\nonumber\\
&&+\gamma_\perp(2\sigma\theta_t\sigma^* - \sigma^*\sigma\theta_t - \theta_t\sigma^*\sigma)dt,\nonumber\\
H&=&\Delta_ca^*a+\Delta_a\sigma^*\sigma+ig_0(a^*\sigma-a\sigma^*)\nonumber\\
&&+i{\cal E}(a^*-a).\label{eq:ucme}
\end{eqnarray}
Here $\kappa$ is the cavity field decay rate, $a$ is the field annihilation operator, $\sigma$ is the atomic dipole (lowering) operator, $\Delta_c=\omega_c-\omega_l$ and $\Delta_a=\omega_a-\omega_l$ are detunings of the cavity and atomic resonance frequencies, and ${\cal E}$ represents the complex amplitude of the coherent driving field. With a view towards matching the usual convention for MBE's, we apply the scalings
\begin{equation}
x\leftarrow \frac{x}{\sqrt{n_0}},\quad p\leftarrow -\frac{p}{\sqrt{2}},\quad D\leftarrow -D,\quad t\leftarrow \gamma_\perp t,
\end{equation}
where $n_0=\gamma_\perp^2/2g_0^2$ is the critical photon number.

After some rather tedious calculations, and adopting the conventional parameter definitions
%\begin{equation}
%$\Delta=\frac{\Delta_a}{\gamma_\perp},\quad \Theta=\frac{\Delta_c}{\kappa},\quad %k=\frac{\kappa}{\gamma_\perp},\nonumber
$\Delta=\Delta_a/\gamma_\perp$, $\Theta=\Delta_c/\kappa$, $k=\kappa/\gamma_\perp$,
%\end{equation}
%\begin{equation}
%C=\frac{g_0^2}{2\kappa\gamma_\perp},\quad y=\frac{\cal E}{\kappa\sqrt{n_0}}$,
$C=g_0^2/2\kappa\gamma_\perp$, $y={\cal E}/\kappa\sqrt{n_0}$,
%\end{equation}
we arrive at the dimensionless equations
\begin{equation}
\frac{dn}{dt}=0,\quad \frac{dD}{dt}=-2(D-1)-2(p_rx_r+p_ix_i),\label{eq:mbef1}
\end{equation}
\begin{equation}
\frac{dp_r}{dt}=-p_r+\Delta p_i+Dx_r,\quad \frac{dp_i}{dt}=-p_i-\Delta p_r+Dx_i,\label{eq:mbef2}
\end{equation}
\begin{eqnarray}
\frac{dx_r}{dt}&=&-k\left\{x_r-\Theta x_i-{\rm Re}[y]+2Cp_rF\right\},\nonumber\\
\frac{dx_i}{dt}&=&-k\left\{x_i-\Theta x_r-{\rm Im}[y]+2Cp_iF\right\},\nonumber\\
F&=&\frac{2}{1+2p_r^2+2p_i^2+D^2}.\label{eq:mbef3}
\end{eqnarray}
We have invoked the fact that $n$ is constant and set $n=1$ in the equations above (as is required for normalization of the projected density operator $\rho_t$). Looking at the resulting equations, we see that we would exactly recover the usual MBE's by setting $F\rightarrow 1$. Given the definitions and scalings used above, we can write $F=2/(1+\vert\vec{S}\vert^2)$, where $\vec{S}$ is the atomic Bloch vector. It follows that for atomic pure states $F\rightarrow 1$ and we can justify the usual MBE's if we have reason to believe that this purity is nearly preserved by the dynamics ({\it e.g.}, under very weak excitation). Likewise we see that $1\le F\le 2$, but comparison of Eqs.~(\ref{eq:mbef1})-(\ref{eq:mbef3}) with the usual MBE's in a standard setting such as absorptive bistability shows that even this factor can lead to significant corrections (see below).

Turning now to the projection of conditional master equations, we first note that terms associated with conditioning upon homodyne/heterodyne detection of the cavity output field vanish upon projection. We therefore begin with the quantum filter for homodyne detection of the atomic fluorescence \cite{Bout04}, written in linear Stratonovich form as appropriate for the projection method (as discussed in \cite{vanH05}):
\begin{eqnarray}
d\theta_t&=&-i[H,\theta_t]dt+\kappa(2a\theta_ta^*-a^*a\theta_t-\theta_ta^*a)dt
\nonumber\\
&&-\gamma_\perp(\sigma^*\sigma\theta_t+\theta_t\sigma^*\sigma)dt\nonumber\\&&
+\sqrt{2\gamma_\perp}(\sigma\theta_t+\theta_t\sigma^*)\circ dy_t.\label{eq:qfho}
\end{eqnarray}
Here $dy_t$ represents the measured photocurrent that drives the filter.
%, for which we can write
%\begin{equation}
%dy_t=d\tilde{W}_t+\sqrt{2\gamma_\perp}\frac{{\rm Tr}[\sigma_x\theta_t]}{{\rm Tr}[\theta_t]},
%\end{equation}
%where $\tilde{W}_t$ is a Brownian motion. A
Here we are not yet scaling time by $\gamma_\perp$. The parameters and the Hamiltonian are as in the unconditional master equation~(\ref{eq:ucme}).

Projection leads to equations including a nontrivial evolution of the normalization,
%\begin{equation}
$dn=\gamma_\perp(D-n)dt-2\sqrt{\gamma_\perp}p_r\circ dy_t$,
%\end{equation}
which is not surprising since we started from an unnormalized model. We therefore transform to normalized variables by defining
%\begin{equation}
$\tilde{p}_r\equiv p_r/n$, $\tilde{p}_i\equiv p_i/n$, $\tilde{D}\equiv D/n$.
%\end{equation}
Converting the resulting equations to It\^{o} form we obtain the projected filter
\begin{eqnarray}
d\tilde{p}_r&=&\gamma_\perp(-3\tilde{p}_r+\Delta\tilde{p}_i+2\tilde{p}_r\tilde{D}+\tilde{D}x_r+4\tilde{p}_r^3)dt
\nonumber\\&&+\sqrt{\gamma_\perp}(2\tilde{p}_r^2+\tilde{D}-1)dy_t,\nonumber\\
d\tilde{p}_i&=&\gamma_\perp(-\Delta\tilde{p}_r-\tilde{p}_i+\tilde{D}x_i+4\tilde{p}_r^2\tilde{p}_i)dt
\nonumber\\&&+2\sqrt{\gamma_\perp}\tilde{p}_r\tilde{p}_idy_t,\nonumber\\
d\tilde{D}&=&\gamma_\perp(2-2\tilde{D}-2\tilde{p}_rx_r-2\tilde{p}_ix_i-4\tilde{p}_r^2+4\tilde{p}_r^2\tilde{D})dt
\nonumber\\&&-2\sqrt{\gamma_\perp}(\tilde{p}_r-\tilde{p}_r\tilde{D})dy_t,\nonumber\\
\frac{dx_r}{dt}&=&-k\gamma_\perp(x_r-\Theta x_i-{\rm Re}[y]+2C\tilde{p}_rF),\nonumber\\
\frac{dx_i}{dt}&=&-k\gamma_\perp(x_i+\Theta x_r-{\rm Im}[y]+2C\tilde{p}_iF).\label{eq:pfho}
\end{eqnarray}

As it is our aim to obtain stochastic equations that could be used for Monte Carlo simulation of semi-classical cavity QED dynamics, we now make the substitution
\begin{equation}
dy_t\rightarrow dW_t-2\sqrt{\gamma_\perp}\tilde{p}_rdt,\label{eq:approx}
\end{equation}
where $dW_t$ is an It\^{o} increment and $-2\sqrt{\gamma_\perp}\tilde{p}_r$ represents an approximation within the semi-classical state space of $\sqrt{2\gamma_\perp}{\rm Tr}[\sigma_x\theta_t]/{\rm Tr}[\theta_t]$, which is the expected value of the measured signal $dy_t$. Making this substitution, and scaling time by $t\leftarrow\gamma_\perp t$, $dW_t\leftarrow\sqrt{\gamma_\perp}dW_t$, we obtain
\begin{eqnarray}
d\tilde{p}_r&=&(-\tilde{p}_r+\Delta\tilde{p}_i+\tilde{D}x_r)dt%\nonumber\\&&
+(2\tilde{p}_r^2+\tilde{D}-1)dW_t,\nonumber\\
d\tilde{p}_i&=&(-\tilde{p}_i-\Delta\tilde{p}_r+\tilde{D}x_i)dt%\nonumber\\&&
+2\tilde{p}_r\tilde{p}_idW_t,\nonumber\\
d\tilde{D}&=&(2-2\tilde{D}-2\tilde{p}_rx_r-2\tilde{p}_ix_i)dt%\nonumber\\&&
+2\tilde{p}_r(\tilde{D}-1)dW_t,\nonumber\\
\frac{dx_r}{dt}&=&-k(x_r-\Theta x_i-{\rm Re}[y]+2C\tilde{p}_rF),\nonumber\\
\frac{dx_i}{dt}&=&-k(x_i+\Theta x_r-{\rm Im}[y]+2C\tilde{p}_iF).\label{eq:ssho}
\end{eqnarray}
Using these equations it can be shown that
\begin{equation}
d(2\tilde{p}_r^2+2\tilde{p}_i^2+\tilde{D}^2)\propto(2\tilde{p}_r^2+2\tilde{p}_i^2+\tilde{D}^2-1)%\nonumber\\
%&&\times [(2-4\tilde{p}_r^2)dt+4\tilde{p}_rdW_t],
\end{equation}
and thus vanishes under the initial condition $2\tilde{p}_r^2+2\tilde{p}_i^2+\tilde{D}^2=1$. Hence for an initial atomic pure state we can simplify the field evolution equations to
\begin{eqnarray}
\frac{dx_r}{dt}&=&-k(x_r-\Theta x_i-y+2C\tilde{p}_r),\nonumber\\
\frac{dx_i}{dt}&=&-k(x_i+\Theta x_r+2C\tilde{p}_i).
\end{eqnarray}

Following an analogous procedure for a linear Stratonovich quantum filter for heterodyne detection of the atomic fluorescence,
\begin{eqnarray}
d\theta_t&=&-i[H,\theta_t]dt+\kappa(2a\theta_ta^*-a^*a\theta_t-\theta_ta^*a)dt\nonumber\\
&&-\gamma_\perp(\sigma^*\sigma\theta_t+\theta_t\sigma^*\sigma)dt\nonumber\\
&&+\sqrt{2\gamma_\perp}(\sigma\theta_t+\theta_t\sigma^*)\circ{\rm Re}[dy_t]\nonumber\\
&&+i\sqrt{2\gamma_\perp}(\sigma\theta_t-\theta_t\sigma^*)\circ{\rm Im}[dy_t],\label{eq:qfhe}
\end{eqnarray}
where $dy_t$ now represents a complex photocurrent as in \cite{Wise93,Goet94}, we obtain
\begin{eqnarray}
d\tilde{p}_r&=&(-\tilde{p}_r+\Delta\tilde{p}_i+\tilde{D}x_r)dt\nonumber\\&&
+\frac{1}{\sqrt{2}}(2\tilde{p}_r^2+\tilde{D}-1)dW^r_t-\sqrt{2}\tilde{p}_r\tilde{p}_idW^i_t,\nonumber\\
d\tilde{p}_i&=&(-\tilde{p}_i-\Delta\tilde{p}_r+\tilde{D}x_i)dt\nonumber\\&&
+\sqrt{2}\tilde{p}_r\tilde{p}_idW^r_t-\frac{1}{\sqrt{2}}(2\tilde{p}_i^2+\tilde{D}-1)dW^i_t,\nonumber\\
d\tilde{D}&=&(2-2\tilde{D}-2\tilde{p}_rx_r-2\tilde{p}_ix_i)dt\nonumber\\&&
+\sqrt{2}\tilde{p}_r(\tilde{D}-1)dW^r_t-\sqrt{2}\tilde{p}_i(\tilde{D}-1)dW^i_t,\nonumber\\
\frac{dx_r}{dt}&=&-k(x_r-\Theta x_i-{\rm Re}[y]+2C\tilde{p}_rF),\nonumber\\
\frac{dx_i}{dt}&=&-k(x_i+\Theta x_r-{\rm Im}[y]+2C\tilde{p}_iF).\label{eq:sshe}
\end{eqnarray}
Here $dW^r_t$ and $dW^i_t$ are independent It\^{o} increments.

\begin{figure}[t!]
\includegraphics[width=0.485\textwidth]{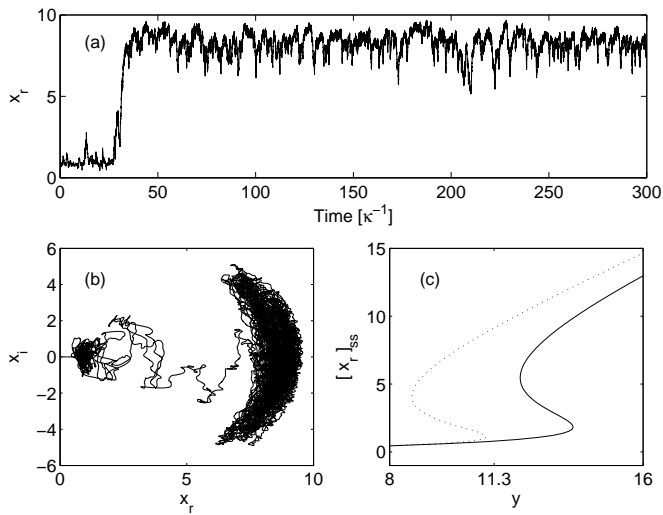}
\caption{\label{fig:sims} (a) Simulated trajectory of $x_r$ from model~(\ref{eq:sshe}) with parameters of absorptive bistability. (b) Joint distribution of $x_r$ and $x_i$ from the same simulation. (c) Comparison of equilibrium solutions according to the usual MBE's (dashed) and our projected equations~(\ref{eq:mbef1})-(\ref{eq:mbef3}).}
\vspace{-0.1in}
\end{figure}

Summarizing the analytic results, we have shown that projection according Eq.~(\ref{eq:proj}) of quantum master equations onto the semi-classical manifold of states specified by Eq.~(\ref{eq:manif}) leads to Maxwell-Bloch-type equations for single-atom cavity QED. Starting from the unconditional master equation~(\ref{eq:ucme}) we obtain the model~(\ref{eq:mbef1})-(\ref{eq:mbef3}). From the quantum filter for homodyne detection of the atomic fluorescence, Eq.~(\ref{eq:qfho}), we obtain the projected filter~(\ref{eq:pfho}) and then use the approximation~(\ref{eq:approx}) to obtain the stochastic simulation model~(\ref{eq:ssho}). From the heterodyne quantum filter, Eq.~(\ref{eq:qfhe}), we obtain by an analogous procedure the stochastic simulation model~(\ref{eq:sshe}). Conditioning terms associated with homo/heterodyne detection of the cavity output field disappear upon projection.

We turn finally to a brief examination of the behavior of our heterodyne model~(\ref{eq:sshe}) in a parameter regime for which the usual MBE's exhibit absorptive bistability. Fig.~\ref{fig:sims}(a) shows a representative Monte Carlo trajectory of $x_r$ for the parameters ($C=10$, $\Theta=\Delta=0$, $k=0.1$ and $y=11.3$), which are the same as were used in numerical simulations of full quantum models in Figs.~2~and~3 of \cite{Arme06}. Our simulations show both transient localization at the equilibrium point with $x_r\approx 1$ and stochastic jumps to a higher-excitation state with $x_r\approx 8$, but no jumps back down. Previously the higher-excitation state observed in numerical studies of quantum models of absorptive bistability \cite{Sava88,Arme06} has been associated with the upper branch of equilibrium solutions to the MBE's. However, as shown in Fig.~\ref{fig:sims}(c), the semi-classical equations we obtained by projection of the unconditional master equation have a unique equilibrium point with $x_r\approx 1$ for the current parameters. Our simulations of model~(\ref{eq:sshe}) indicate that for our parameters the high-excitation ``state'' is not well-localized in phase space, as seen in Fig.~\ref{fig:sims}(b). This plot of $x_r$ versus $x_i$ shows a distribution of intracavity field amplitude similar to the steady-state Q-function obtained from the unconditional master equation (see Fig.~2 of \cite{Arme06}). It may therefore prove enlightening to study the trapping region that apparently exists in the semi-classical phase portrait near $x_r\approx 8$, as the intermittency here induced by quantum fluctuations is seen to be more general than the previously-known phenomenon of jumping between mean-field equilibria \cite{Sava88,Alsi91,Mabu98,Arme06}.

\begin{acknowledgments}
The author would like to thank Ramon~van~Handel for invaluable assistance.
\end{acknowledgments}

\end{document}